\documentclass[sigconf,review]{acmart}

\renewcommand\footnotetextcopyrightpermission[1]{}
\setcopyright{acmcopyright}
\copyrightyear{2020}
\acmYear{2020}
\setcopyright{acmcopyright}\acmConference[CPSS '20]{Proceedings of the 6th ACM Cyber-Physical System Security Workshop}{October 6, 2020}{Taipei, Taiwan}
\acmBooktitle{Proceedings of the 6th ACM Cyber-Physical System Security Workshop (CPSS '20), October 6, 2020, Taipei, Taiwan}
\acmPrice{15.00}
\acmDOI{10.1145/3384941.3409588}
\acmISBN{978-1-4503-7608-2/20/10}

\usepackage{soul}
\usepackage{amsmath,amsfonts}
\usepackage{algorithmic}
\usepackage[ruled,vlined ]{algorithm2e}
\usepackage{multirow}
\usepackage{array}
\usepackage{tabularx}
\usepackage{graphicx}
\usepackage{textcomp}
\usepackage{xcolor}
\usepackage{psfrag}
\usepackage{mathtools}
\usepackage{xspace}
\def\BibTeX{{\rm B\kern-.05em{\sc i\kern-.025em b}\kern-.08em
    T\kern-.1667em\lower.7ex\hbox{E}\kern-.125emX}}

\newcounter{challenge}

\newcounter{recommendation}

\newcounter{myposition}

\usepackage{balance}

\usepackage{listings}
\usepackage{xcolor}
 
\definecolor{codegreen}{rgb}{0,0.6,0}
\definecolor{codegray}{rgb}{0.5,0.5,0.5}
\definecolor{codepurple}{rgb}{0.58,0,0.82}
\definecolor{backcolour}{rgb}{0.95,0.95,0.92}
 
\lstdefinestyle{mystyle}{
    backgroundcolor=\color{backcolour},   
    commentstyle=\color{codegreen},
    keywordstyle=\color{magenta},
    numberstyle=\tiny\color{codegray},
    stringstyle=\color{codepurple},
    basicstyle=\ttfamily\footnotesize,
    breakatwhitespace=false,         
    breaklines=true,                 
    captionpos=b,                    
    keepspaces=true,                 
    numbers=left,                    
    numbersep=5pt,                  
    showspaces=false,                
    showstringspaces=false,
    showtabs=false,                  
    tabsize=2
}
 
\def\adityaIgnore#1{}

\begin{document}
\title{Attack Rules: An Adversarial Approach to Generate Attacks for Industrial Control Systems using Machine Learning}
%prof and Mujeeb, kindly verify the title 

\fancyhead{}

%\title{Challenges in using Machine Learning for Detecting Cyber Attacks  in Industrial Control Systems}

%{\footnotesize \textsuperscript{*}Note: Sub-titles are not captured in Xplore and
%should not be used}
%\thanks{Identify applicable funding agency here. If none, delete this.}
%}

% \author{\IEEEauthorblockN{Chuadhry Mujeeb Ahmed\IEEEauthorrefmark{1}, Gauthama Raman M R\IEEEauthorrefmark{2}, Aditya Mathur\IEEEauthorrefmark{3}}
% \IEEEauthorblockA{iTrust, Center for Research in Cyber Security\\
% Singapore University of Technology and Design\\
% Singapore\\
% \IEEEauthorrefmark{1}chuadhry@mymail.sutd.edu.sg,
% \IEEEauthorrefmark{2}gauthama\_mani@sutd.edu.sg,
% \IEEEauthorrefmark{3}aditya\_mathur@sutd.edu.sg}
% }

\author{Muhammad Azmi Umer}
\affiliation{%
  \institution{DHA Suffa University}
  \institution{Karachi Institute of Economics and Technology}
%   \streetaddress{1 Th{\o}rv{\"a}ld Circle}
%   \city{Hekla}
%   \country{Iceland}
}
\email{azmi.umer@dsu.edu.pk}

\author{Chuadhry Mujeeb Ahmed}
\affiliation{%
  \institution{University of Strathclyde}
%   \streetaddress{1 Th{\o}rv{\"a}ld Circle}
%   \city{Hekla}
%   \country{Iceland}
}
\email{mujeeb.ahmed@strath.ac.uk}

\author{Muhammad Taha Jilani}
\affiliation{%
  \institution{Karachi Institute of Economics and Technology}
%   \streetaddress{1 Th{\o}rv{\"a}ld Circle}
%   \city{Hekla}
%   \country{Iceland}
}
\email{m.taha@pafkiet.edu.pk}

\author{Aditya P. Mathur}
\affiliation{%
  \institution{Singapore University of Technology and Design}
%   \streetaddress{1 Th{\o}rv{\"a}ld Circle}
%   \city{Hekla}
%   \country{Iceland}
}
\email{aditya\_mathur@sutd.edu.sg}

\begin{abstract}
\adityaIgnore{Machine learning has found applications in the security domain, especially for attack detection. At the same time, adversarial learning has received a lot of attention from the research community in the cybersecurity domain.} Adversarial learning is used to  test the robustness of machine learning algorithms under attack and create attacks that deceive the  anomaly detection methods in Industrial Control System (ICS).\adityaIgnore{ However,  little attention is paid to the exhaustive search of the attack space.} Given that security assessment of an ICS  demands that an exhaustive  set of possible attack patterns is studied,  in this work, we propose an association rule mining-based attack generation technique. The technique has  been implemented using data from a Secure Water Treatment plant. The proposed technique was able to generate more than 300,000 attack patterns constituting a vast majority of new attack vectors which were not seen before. Automatically generated attacks improve our  understanding of the potential attacks and  enable the  design of  robust attack detection techniques. 

\end{abstract}


\maketitle

% \begin{IEEEkeywords}
% Anomaly detection; Critical Infrastructure; Industrial Control Systems; Machine Learning; System Identification; Water Treatment Plant.
% \end{IEEEkeywords}

%\input{intro.tex}
%===============

\section{Introduction}
% Introduction to the CPS and the attack detection problem

Most of the modern critical infrastructures (CI) are controlled by Industrial Control Systems (ICS). Examples of such systems are the electric power grid and water treatment plants, where an ICS controls the physical process in a CI. Computing and communication elements such as Programmable Logic Controllers (PLCs) and  Supervisory Control and Data Acquisition systems (SCADA), and communication networks are the important technologies used to realize the control. PLCs control the physical process via actuators based on the sensor measurements. The SCADA workstations are used to exert high-level control over the PLCs, and the process, and provide a view into the current process state~\cite{ScanningTheCycle_AsiaCCS2021}. The automation makes it easy to monitor and control the critical infrastructures but it opens up these critical systems to malicious entities as evident from several widely reported successful attempts such as those reported in,\cite{weinbergerStuxnet,ukraineBlackout,germanSteelMill}. It is observed that while air-gaping a system might be a means to consider securing a CI, it does not guarantee to keep attackers from gaining access to the ICS and launch successful attacks.

Due to the increase in attacks on ICS, a lot of attention is given to the development of security mechanisms to prevent, contain, and react to cyber-attacks~\cite{challenges_SnP_ahmed2020}. This led to research in two major directions, 1) rigorous security testing of ICS and 2) design of robust intrusion detection techniques. The outcome of those efforts either relied on the design of these complicated systems or the data from the CI. On the security analysis front, one approach was to come up with realistic threat models, accurately modeling the capabilities and intentions of an attacker specifically applied to an ICS~\cite{sridhar_compsac_2016,Rocchetto2016}. These earlier works provide interesting frameworks to model attacks in an ICS based on an attacker's capabilities and intentions but require design knowledge and domain expertise in the specific process. In this work, we propose a data-based automatic attack generation technique to drive an ICS to an unsafe state, this type of resource would prove to be a key technique to cover as many bases as possible. 

Other efforts towards generating the anomalous data rely on the normal process data to come up with examples not present in the normal data~\cite{pham2014generating_attacks}. However, data generated this way might be anomalous but the particular anomaly doesn't need to have any strategy to cause any damage and labeled as the attack data. Studies looked into generating more samples of attack data than really present, e.g., SMOTE~\cite{chawla2002smote}, but those do not add any new attack patterns rather interpolate the existing ones. Therefore, we take up this study to design an attack generator that can learn from the attack goals of the attacks designed by domain experts. Association Rule Mining (ARM) was used to find attack scenarios autonomously using the attack samples. Earlier manually generated attacks~\cite{sridhar_compsac_2016} were crafted by a human expert to take the system to an unsafe state. Given those sets of actions and knowing that these will take a system into an unsafe state, we can learn more about such patterns that were not explored before. This leads to, 1) discovering attacks that were not seen before, 2) analyze the impact of those attacks. We plan to exhaust the possibilities to bring a physical process in an unsafe state and figure out sensor/actuator combinations to be manipulated. 

% \textcolor{red}{@Azmi: It would be interesting to have some sort of metric like this, a particular unsafe state, e.g., tank overflow and number of ways (patterns mined) this can be achieved and/or number of devices to be manipulated to achieve this or number of attributes....Or support as in duration of attacks.} 

% \skipnoindent{\em Organization}: The remainder of this paper is organized as follows. Section~\ref{sec:swat} is a brief introduction to a live water treatment plant used extensively by the authors for testing anomaly detectors derived using process data. Terminology related to anomaly detectors is explained in Section~\ref{sec:anomalyDetectors}. Challenges in the design of anomaly detectors using plant data are enumerated and explained in Section~\ref{sec:challenges}. Research directions aimed at the development of methods to overcome the challenges are summarized in Section\ref{sec:futureWork}.

% \textcolor{red}{A lot of interesting stuff and paragraphs we can reuse from this paper of Ensuk \cite{Eunsuk_Sridhar_Modelbased}: following from Eunsuk paper but we can say something similar when we get the results}
%https://dspace.mit.edu/bitstream/handle/1721.1/114444/Model-based\%20security.pdf?sequence=1&isAllowed=y}

In this work, a water treatment testbed is used as a case study and our proposed technique is applied to generate attack patterns. As expected we obtained 329069 attack patterns that were impossible to find manually. SWaT testbed engineers spent a lot of effort and ran the plant continuously for 4 days to come up with 36 attack patterns across the plant~\cite{sridhar_dataset_paper}. This is considered a remarkable effort, given the limited research facilities and lack of data availability especially the attack data. However, for data-hungry machine learning algorithms, this is still a limited amount of data and lacks completeness if someone wants to design a supervised learning algorithm for attack detection. The generated attack patterns are verified to be attacks using the well-known \emph{process invariants} technique~\cite{UMER_Azmi_2020} that mined the process invariant from the normal operational data of the plant.

\noindent \emph{Contributions:} Contributions of this work are twofold. 1) Security analysis of the physical process by exploring the attack patterns. This is to explore the combination of possible sensor's and actuator's states to lead to unsafe operation. 2) Providing the data for improving the design of intrusion detection systems. Data is the first requirement for machine learning-based intrusion detection techniques. Normal data could be easily obtained from the physical operation of the process but attack data is rarely available. In cases where the attack data is collected, the attacks generated by human experts are not exhaustive. However, we have tried to provide all the combinations that can lead to successful attacks.

\begin{figure}
    \centering
    \includegraphics[scale=0.25]{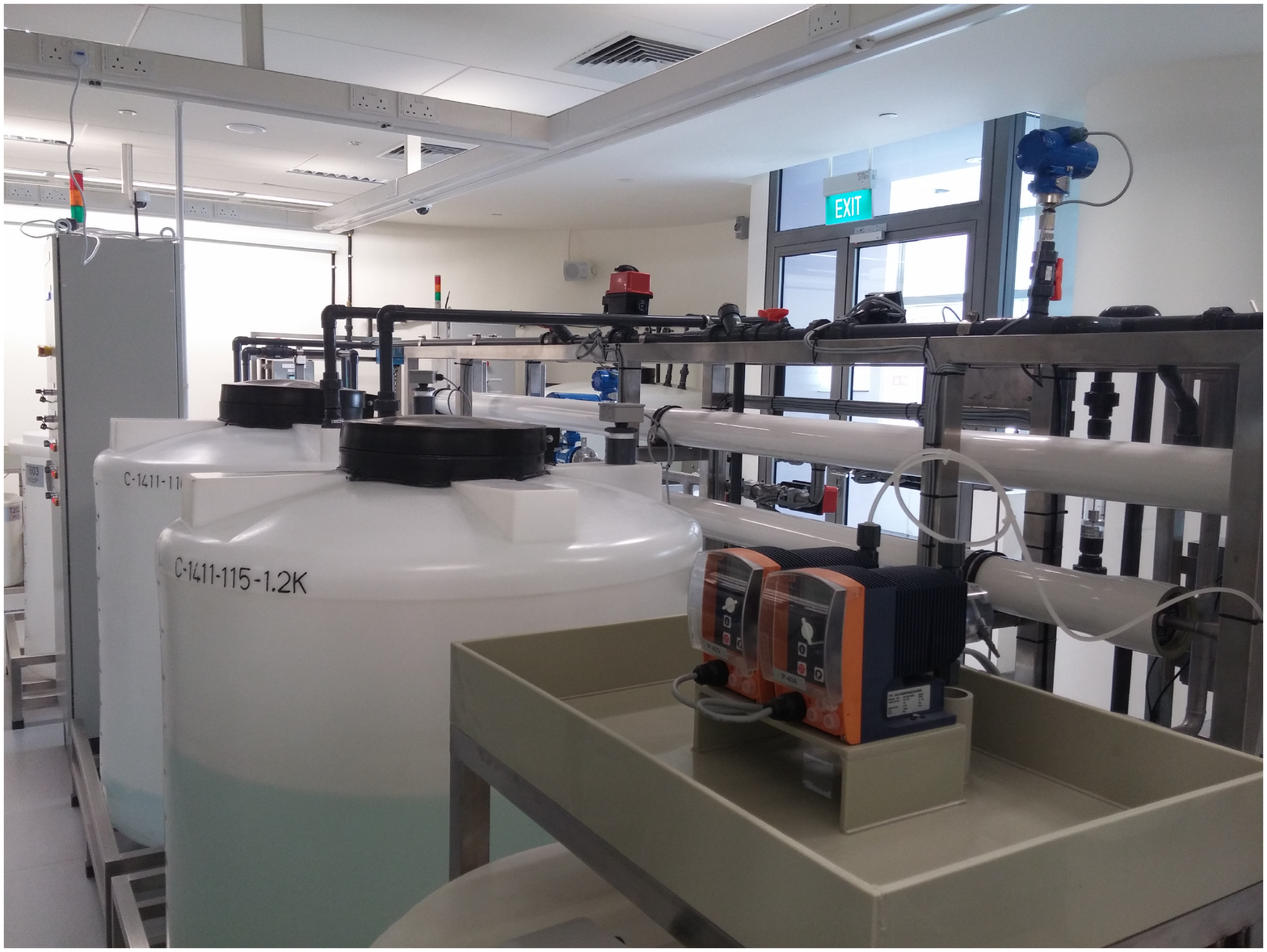}
    \caption{Case Study: SWaT Testbed}
    \label{fig:SWaT_Testbed_Pic}
\end{figure}

\section{Background}

\subsection{SWAT: A live water treatment plant}
\label{sec:swat}
% an explanation on the task of attack detection in  a water system is given here

The Secure Water Treatment (SWaT) is an industrial replica of a water treatment plant which is available at Singapore University of Technology and Design (SUTD)\,\cite{mathurTippenhauer}. It has been used by various researchers to validate their defense mechanisms for ICS\,\cite{swatDataset}. A pictorial view of the SWaT plant is shown in Figure~\ref{fig:SWaT_Testbed_Pic}. It has the capacity to produce 5\,gallons/minute of treated water. Water is treated first using ultrafiltration and then followed by reverse osmosis. SWaT is a  distributed control system consisting of six stages. Each stage is installed with a set of sensors and actuators. Sensors are used to measure the quantity of water like the level in a tank, flow, and pressure. They are also used to measure the quality of water like pH, oxidation-reduction potential, and conductivity. Motorized valves and electric pumps serve as actuators.  
 
 Stage\,1 treats the raw water. Chemical dosing is performed in stage\,2 that depends on the measurements from the water quality sensors. Ultrafiltration occurs in stage\,3. Dechlorination is performed in stage\,4 before it is passed to the reverse osmosis units in stage\,5. The treated water is distributed in stage\,6 and it also performs the backwash by cleaning the ultrafiltration unit. Level\,0 network is used for data communication between sensors/ actuators and PLCs. Inter PLCs communication is done over a level\,1 network.

\begin{figure}
    \centering
    \includegraphics[scale=0.3]{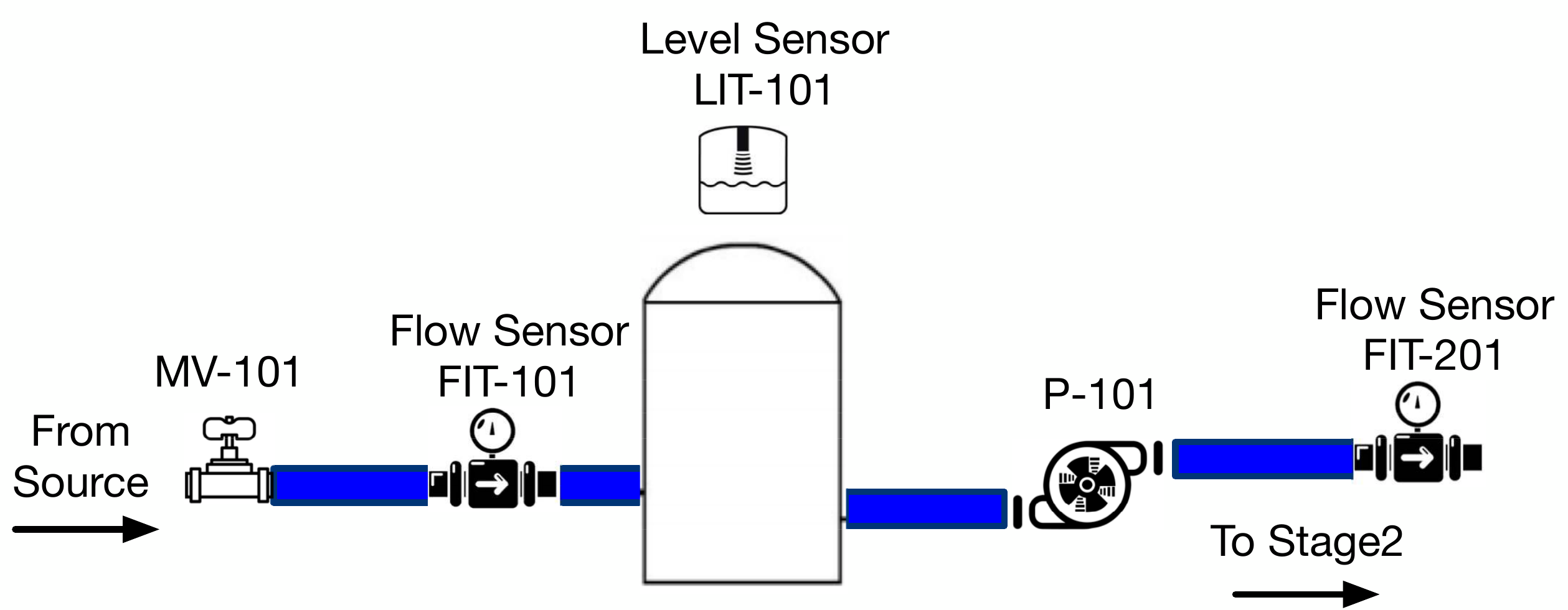}
    \caption{Stage~1 of the SWaT Testbed.}
    \label{fig:Stage1_SWaT}
\end{figure}

\subsection{Motivating Example}

Figure~\ref{fig:Stage1_SWaT} shows stage~1 of the SWaT testbed. It consists of a raw water storage tank, a level sensor (LIT-101) on top of the tank, an inlet motorized valve (MV-101), a flow meter (FIT-101) on the inlet pipe, a pump (P-101) on the outlet, a standby pump (P-102 not shown in the figure), a flow sensor (FIT-201) and a motorized valve (MV-201) on the outlet of the tank and on the stage~2 of the SWaT. Now, having seen the critical components let's see how does this process is controlled by a PLC. The purpose of the tank is to store the raw water to be supplied to subsequent stages for the purification process. Listing~1 shows a snippet of a part of the control logic in the PLC.

\vspace{0.2cm}

\begin{lstlisting}[language=Python, caption=Control Logic Example in Stage1 of SWaT]
# LIT-101 is level sensor in tank1
# MV-101 is the inlet valve controlled by a PLC depending on LIT101 measurements. 
if LIT-101_value == Low_Level:
    MV-101 = Open
elif LIT-101_value == High_Level:
    MV-101 = Close
 \end{lstlisting}
 
 There are designed operational limits for the tank to hold the water, i.e., the Low\_Level is the lowest limit beyond which stage~1 would not be able to meet the water requirement of the subsequent stages and High\_Level is the upper limit beyond which the tank would overflow. According to the control logic shown in Listing~1 when the tank is at a high level, i.e., full then the inlet supply via MV-101 should be cut off to avoid overflow by closing the MV-101. If MV-101 does not close ultimately tank would overflow leading to a flooding situation. Now imagine that an attacker's goal is to overflow the tank, this goal could be achieved in many ways. One way is to force the MV-101 to stay open although the water level is high. This attack scenario compromises only one component MV-101 and succeeds but this attack could be easily detected by an operator monitoring the SCADA system by observing that the level sensor measurements have gone way higher than those should be. Another possible attack could compromise both the level sensor LIT-101 and the MV-101, by keeping the MV-101 open and spoofing the sensor reading within low and high levels. This way simply monitoring level sensor would not raise any alarms and the SCADA system would also fail to report this attack. We could see that by changing more than one component our attack becomes more sophisticated. Stage~1 is quite simple but still has got 7 variables that could be modified in different combinations to create a lot of attack scenarios. Given that the other stages in SWaT and real-world systems are much more complex with sometimes 100s of devices, then it is really hard to generate all possible attacks manually. Therefore, in this work, we have proposed a data-based autonomous technique to generate possible attack patterns.

\section{Attacks Generation Using Association Rule Mining}
\label{sec:arm}

Lack of comprehensive attack scenarios and data is a major concern while working on anomaly detection problems. The key idea of the proposed technique is to use the available attack data~\cite{sridhar_dataset_paper} created by human experts and generate an exhaustive list of possible attack patterns. Attack data on the SWaT testbed is collected for $4$ days while running the plant continuously. Plant engineers and researchers from SUTD came up with a rigorous design of attacks modeling the attacker's intentions, capabilities, and attack goals. An example of one such attack is the tank overflow attack as explained in the previous section. Therefore, an attack model is a tuple composed of sensor/actuator pair it shall attack with a consequence on the real physical process. This attack generation exercise led to the generation of $36$ attack patterns in total. This attack dataset is not exhaustive in any sense but it is really useful for machine learning-based attack/defense techniques. Our proposed idea leverages that dataset and comes up with new attack patterns. The key observation was that each of those $36$ attacks had a goal in mind, e.g., tank overflow and attack is executed by operators with design knowledge to incur the required damage. We model those intentions from the sample attack dataset and tried to come up with all the possible ways in which that goal of an attacker is fulfilled. To capture those rules set we have used the Association Rule Mining (ARM) \cite{agrawal1993mining} for attack generation. ARM is a rule-based unsupervised ML technique. It is very common in traditional market basket analysis. However, it also has important applications in intrusion detection \cite{ahmed2021machine}. In the current study, ARM was used for attack generation on a secure water treatment plant (SWaT) as described in algorithm \ref{alg:in}. Attacks were generated using the attacked data of SWaT \cite{swatDataset}. The following type of attacks was generated using the proposed approach:

\begin{equation}
\label{eq:in}
    X\;\Longrightarrow\; Y
\end{equation}

\begin{figure*}
    \centering
    \includegraphics[scale=0.6]{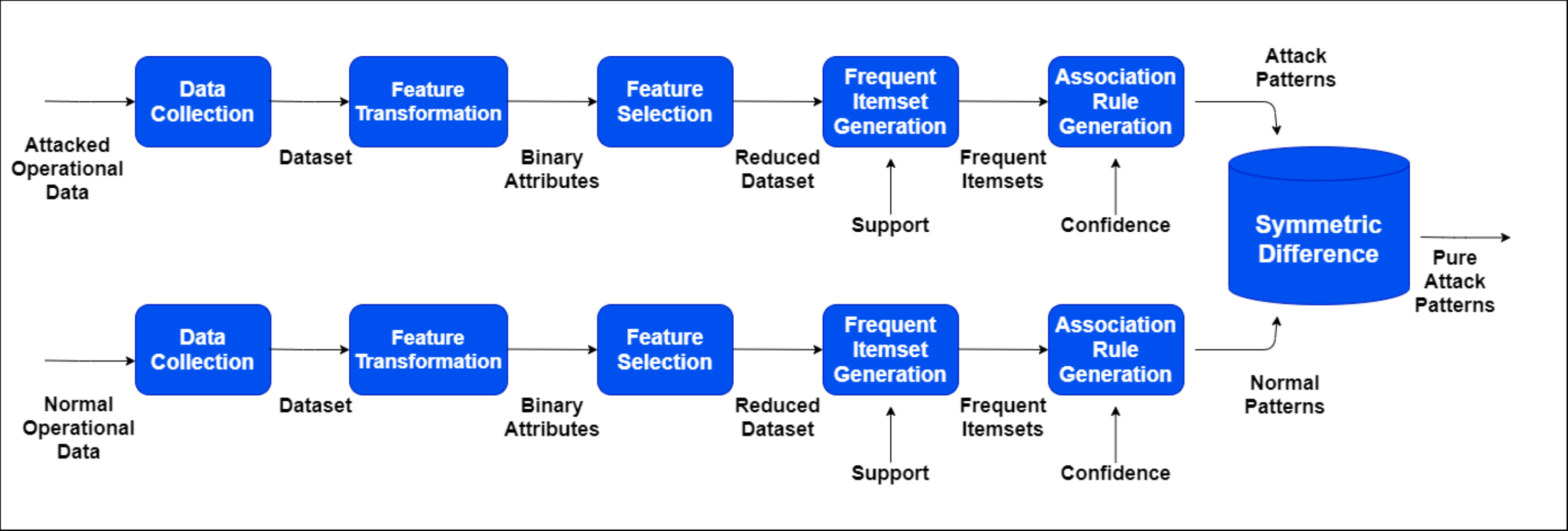}
    \caption{Attacks Generation using ARM}
    \label{fig:genval}
\end{figure*}

There are several algorithms including Apriori, Eclat, and FP-Groth which can be used for ARM. FP-growth was used in the given case study which is available in the Orange-Associate library in Python. ARM works in two phases. In the first phase, it generates frequent itemsets, while in the second phase it generates association rules using the frequent itemsets generated in the first phase. The complete flow of work is described in Figure \ref{fig:genval}.

\subsection{Frequent Itemsets}
An itemset is either a value of a single attribute or values of multiple attributes in a dataset. Itemset that qualifies the minimum support threshold is considered as a frequent itemset.

\subsubsection{Support:} Let there is an itemset 'I' in dataset 'D'. The support for an itemset 'I' in 'D' can be calculated by counting the number of transactions or rows 'r' of 'D' which contains the 'I'. 

\begin{equation}
\textstyle{S}(I) = \frac{|r \in D; I \in r|} {|D|}
\end{equation}

\subsection{Association Rules}
Frequent itemsets are used to generate association rules which are described in equation \ref{eq:in}. Only rules which qualify the minimum confidence threshold are considered as association rules. 

\subsubsection{Confidence:} The rule described in equation \ref{eq:in} can be divided into two parts. One is 'X' which is placed at the left side of $\Longrightarrow$ and called as antecedent, while one is 'Y' which is placed at the right side of $\Longrightarrow$ and called as consequent. Confidence of any rule is calculated using the combined support of antecedent and consequent, and the antecedent alone.

\begin{equation}
\textstyle{C}(X \;\Longrightarrow\; Y) = \frac{S(X \cup Y)} {S(X)}
\end{equation}

\begin{algorithm}
\caption{Attack Generation using Association Rule Mining}
\label{alg:in}
\begin{algorithmic}[1]
\STATE Data collection by capturing the network packets

\STATE Decoding of network packets to store state information of sensors and actuators.

\STATE Send the state information to the historian.

\STATE Feature/ Attribute Transformation followed by Feature Selection on collected data.

\STATE Frequent itemsets generation using the transformed data.

\STATE Generation of association rules from the frequent itemsets.

\STATE Validation of attacks (association rules) against the normal patterns. 

\end{algorithmic}
\end{algorithm}

\subsection{Feature Transformation and Selection to Generate Attack Patterns}
SWaTs' dataset consists of 51 attributes consist of different sensors and actuators. They are binary (e.g. pumps), ternary (e.g. motorized valves), and real-valued (e.g. level sensors) attributes. However, ARM works only on binary-valued attributes. Therefore, all other types of attributes need to be transformed into binary-valued attributes. While converting them into binary-valued attributes, some of the attributes become useless, for example after conversion into binary the whole vector of attribute gets converted into a single number i.e. either 0 or 1. This type of data is useless for ARM. Therefore, it needs to be removed from the dataset. There is a total of 15 attributes were selected for attack generation. Another reason for this reduced set of attributes is the baseline validation of attack patterns with normal patterns of SWaT generated in \cite{UMER_Azmi_2020}. We selected the same attributes as done in \cite{UMER_Azmi_2020}.

\begin{table}[htb]
\begin{tabular}{|l|m{18em}|}
\hline
\textbf{Attributes} & \multicolumn{1}{c|}{\textbf{Description}}                                            \\ \hline
\multicolumn{2}{|c|}{\textbf{Flow meters}}                                                                 \\ \hline
FIT-101              & Flow measurement at inlet of T-101.                                                  \\ \hline
FIT-201              & Flow measurement between Process 1 and 2.                                \\ \hline
FIT-301, FIT-601         & Flow measurement in Ultra filtration Process, and backwash respectively.                \\ \hline
\multicolumn{2}{|c|}{\textbf{Motorized valves}}                                                            \\ \hline
MV-101, MV-201          & Motorized Valve at inlet of T-101, and T-301.                                               \\ \hline
MV-301, MV-303          & To control the Ultra filtration-backwash.                                                          \\ \hline
MV-302, MV-304          & To control the flow towards the de-chlorination unit, and Ultra filtration backwash drain respectively \\ \hline
\multicolumn{2}{|c|}{\textbf{Pumps}}                                                                       \\ \hline
P-101                & Pump to supply raw water in Process 2.                                                      \\ \hline
P-203, P-205           & Dosing pumps for HCl , and NaOCl respectively.                          \\ \hline
P-302, P-602           & To supply water from T-301 to T-401, and from T-602 to Ultra filtration unit.                        \\ \hline
\end{tabular}
\caption{Attributes Selected for Attacks Generation.}
\label{tab:inv}
\end{table}

\subsubsection{Transformation of Attributes}

There are some motorized valves in the dataset which have three possible states, i.e., Close (1), Open (2), and Transition (0). This makes them ternary valued attributes. The transition state lasts for less than 10 seconds. To convert these attributes into binary attributes, the transition state (0) was transformed into either Open (2) or Close (1) state. This was done by looking at the corresponding FIT value. If the corresponding FIT Value against the transition state (0) is less than 0.5 then the transition state was converted into Close (1) state else into Open (2) state as shown in Listing~2. The FITs which are the real-valued attribute was transformed into the binary-valued attribute. If its' value is equal to or greater than 0.5 then flow was assumed otherwise not.

\subsubsection{Large Set of Rules}

Defining an optimal value of support is very crucial to the number of rules. If it is too high then the number of rules would be too low, therefore, it is inappropriate to get the actual behavior or patterns in the data. While if support is too low then a very high number of rules gets generated. There was an attribute in SWaT i.e. P602 which remained open only in 3164 transactions of normal SWaT data, the rest of the time it remained close. To get the patterns relevant to P602 in Open state, we need to drop the support level up to 3164/410400, i.e. 0.77\%, where 410400 is the total number of transactions in the dataset.

\begin{lstlisting}[language=Python, caption=Transformation of Motorized Valve into Binary-Valued Attribute]
# MV-101 is the inlet valve controlled by a PLC depending on LIT101 measurements. 
# FIT-101 is use to measure the flow towards Tank T101.
if MV-101 == Open:
    MV-101 = Open
elif MV-101 == Close:
    MV-101 = Close
elif MV-101 == Transition:
    if FIT-101<0.5:
        MV-101 = Close
    else:
        MV-101 = Open 
 \end{lstlisting}
%ML-based anomaly detection approaches normally suffer from zero-day attacks and high false alarms. While applying supervised learning techniques, a lack of attack data creates a bottleneck. Likewise, unsupervised learning approaches suffer from high false alarms. The study reported in this case study has solved both the problems. This is an unsupervised learning approach, so there is no requirement of attack data. Secondly, the proposed approach is capable of detecting a zero-day attack. Because here invariants are generated using benign data of an operational plant. Later these invariants were placed as monitors for anomaly detection. There were no false alarms observed during the operation of the plant. Further, the invariants generated were having antecedent size = 1 to 7. This makes the current approach quite effective against distributed attack detection. A sample of invariants generated using the data-centric approach is described in Table \ref{tab:sinv}. There are 7 types of invariants depending on the size of the antecedent. If antecedent size is 1 then it checks pairwise consistency between different actuators or sensors. While if antecedent size is more than 1 then all the sensors and actuators present in the antecedent must be true to reach the conclusion or consequent. The complete list of invariants is available at \cite{swatDataset}.
 
\section{Evaluation and Discussion}
To validate the generated attack patterns, we replicated the experiments performed in\,\cite{UMER_Azmi_2020} using the data-centric approach. The challenge was that the attack dataset used also had the normal data samples thus making it difficult to identify attacks among the generated patterns based on the defined rules. For this purpose we used the seven days normal operational data from SWaT\,\cite{swatDataset} to generate the physical invariants under the normal operation of the plant. All  attributes were transformed into binary attributes and feature selection was performed to derive the same attributes as done in\,\cite{UMER_Azmi_2020}.  Frequent itemsets were generated using the transformed dataset with 0.7\% support. The frequent itemsets were used to mine the association rules with  100\% confidence level. We compared these normal patterns of SWaT against the attacks generated in the current case study using the same support and confidence thresholds. For this purpose, all the attack patterns of SWaT were converted into a set, say A, where each entry is an attack pattern. Likewise, all  normal patterns of SWaT were converted to a set B where each entry is a normal pattern. We thus obtained  the symmetric difference of sets A and  B  and stored it in set C ($C =  A \setminus  B$ ). Now C contains all the attack patterns that can be differentiated   from the normal patterns.

\begin{figure}
    \centering
    \includegraphics[scale=0.65]{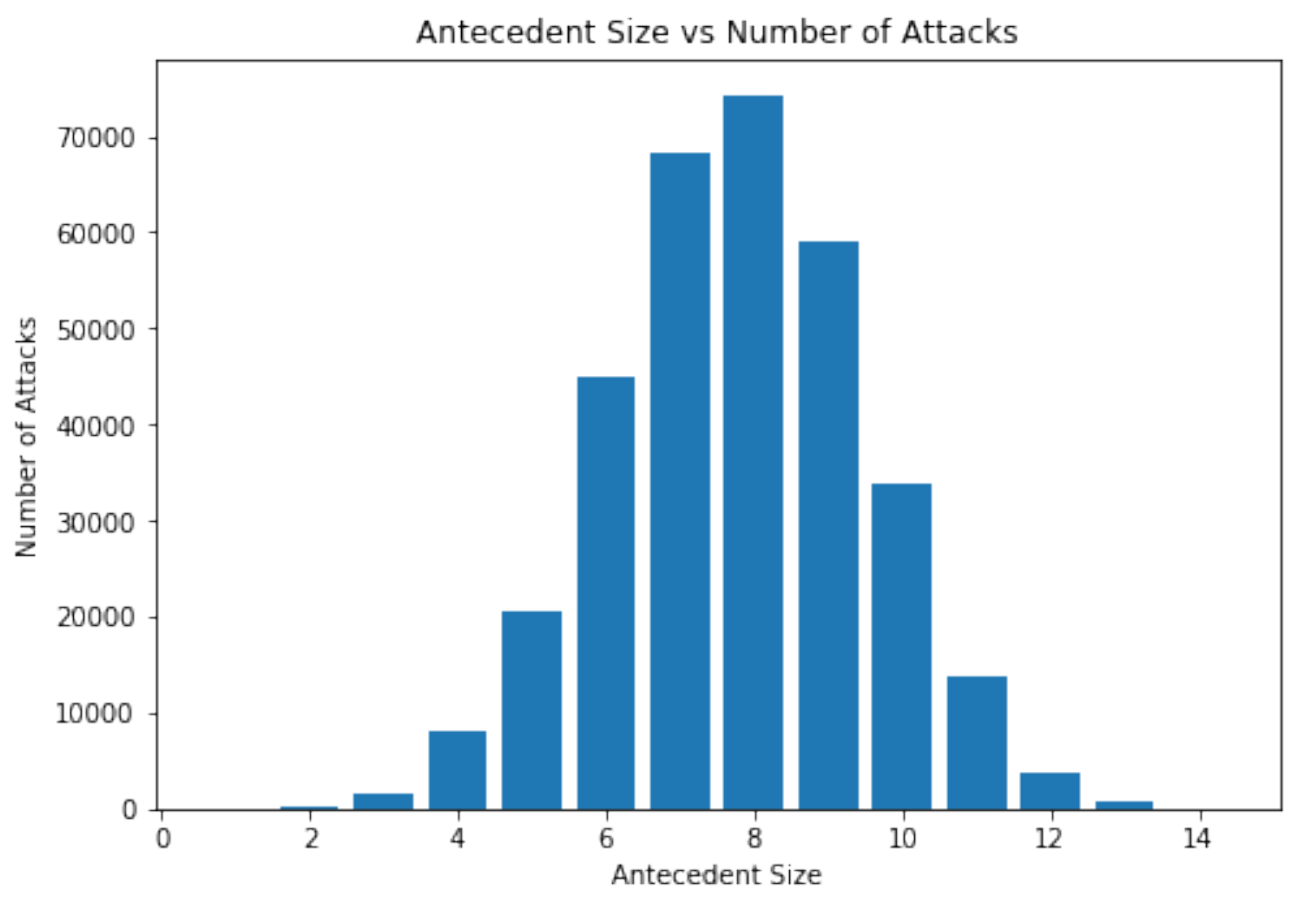}
    \caption{Number of Attacks with respect to Antecedent size}
    \label{fig:antecedent}
\end{figure}

A list of attacks generated using ARM along with sample attack patterns is shown in Table\,\ref{tab:attack_generated}. We were able to generate both local and global attacks. Here ``local attack" refers to   intra-process attack patterns and ``global attack"  refers to inter-process, i.e.,  across 6-stages, attack patterns. We generated attacks from the antecedent size of 2 to the antecedent size of 14 (as a total of 15 sensor/actuator pairs considered in this work). Attack patterns are normally distributed with respect to antecedent size as shown in Figure\,\ref{fig:antecedent}. Figure\,\ref{fig:antecedent} highlights a key result that is ``using a 1 or 2 sensor/actuator pairs, an attacker can launch a small number of attacks but when we increase this number the attack patterns increase exponentially." Results show that by leveraging more sensors and actuators, an attacker can inflict more harm but at some point increasing the devices involved does not result in a successful attack as all the devices involved might not have any physical relationship. With $8$ devices involved, we get the maximum possible attack scenarios.
%However, for too many actuator/sensor pairs, chosen from different stages of SWaT, there would not exist a physical relationship that could be exploited as increasing the number of devices means more independent devices in the set and hence reduced number of attack patterns.
If antecedent $\longrightarrow$ consequent is true then the system should be considered as under attack. These attack patterns are  useful in signature-based intrusion detection. Usually these approaches suffer from a lack of attack patterns in complex physical processes.  Through our approach, we were able to generate more than 300,000 attack patterns for SWaT. We plan to share our results and dataset publicly.

\newcolumntype{C}[1]{>{\centering\let\newline\\\arraybackslash\hspace{0pt}}m{#1}}

\begin{table*}[]
\caption{A sample of attacks generated  using ARM. *The comma (,) is equivalent to Boolean AND. If antecedent $\longrightarrow$ consequent is true then the system should be considered as under attack.}
\label{tab:attack_generated}
\resizebox{17.5cm}{6.25cm}{
\begin{tabular}{|C{2cm}|C{1.5cm}|C{9cm}|C{2cm}|}
\hline
\textbf{Antecedent Size} & \textbf{No. of  Attack Patterns} & \textbf{Antecedent*} & \textbf{Consequent} \\ \hline
2 & 96 & MV304=Close, FIT301\textless{}0.5 & P302=Off \\ \hline
3 & 1554 & MV301=Open, MV302=Close, MV303=Close & MV304=Close \\ \hline
4 & 8133 & P302=Off, P203=Off, MV303=Close, MV304=Open & P602=Off \\ \hline
5 & 20485 & P101=Off, FIT201\textless{}0.5, MV201=Open, MV302=Open, P302=On & MV101=Open \\ \hline
6 & 45006 & FIT101\textless{}0.5, MV101=Close, P203=Off, MV304=Open, P302=Off, FIT601\textless{}0.5 & MV303=Close \\ \hline
7 & 68294 & FIT101\textless{}0.5, MV101=Close, P101=Off, P203=Off, FIT301\textless{}0.5, MV301=Close, P602=Off & MV201=Close \\ \hline
8 & 74348 & MV101=Open, MV201=Open, P203=Off, FIT301\textgreater{}=0.5, MV301=Close, MV302=Open, MV303=Close, MV304=Close & P602=Off \\ \hline
9 & 58936 & MV101=Open, P101=Off, FIT201\textless{}0.5, P203=Off, FIT301\textless{}0.5, MV301=Close, MV302=Open, MV304=Close, P602=Off & MV201=Close \\ \hline
10 & 33903 & FIT101\textgreater{}=0.5, P101=Off, FIT201\textless{}0.5, MV201=Close, P205=Off, MV302=Open, MV303=Close, MV304=Close, P302=Off, FIT601\textless{}0.5 & P602=Off \\ \hline
11 & 13832 & FIT101\textless{}0.5, MV101=Close, MV201=Close, P203=Off, P205=Off, MV301=Close, MV302=Close, MV304=Open, P302=Off, FIT601\textless{}0.5, P602=Off & MV303=Close \\ \hline
12 & 3802 & MOFIT101=Close, MV101=Close, FIT201\textless{}0.5, P203=Off, P205=Off, FIT301\textless{}0.5, MV301=Close, MV303=Close, MV304=Close, P302=Off, FIT601\textless{}0.5, P602=Off & P101=Off \\ \hline
13 & 632 & FIT101\textgreater{}=0.5, MV101=Open, P101=Off, MV201=Close, P203=Off, P205=Off, FIT301\textless{}0.5, MV301=Close, MV302=Open, MV303=Close, P302=Off, FIT601\textless{}0.5, P602=OFF & MV304=Close \\ \hline
14 & 48 & FIT101\textless{}0.5, MV101=Close, P101=Off, FIT201\textless{}0.5, MV201=Close, P203=Off, P205=Off, FIT301\textless{}0.5, MV301=Close, MV303=Close, MV304=Open, P302=Off, FIT601\textless{}0.5, P602=Off & MV302=Close \\ \hline
\end{tabular}%
}
\end{table*}

\section{Related Work}

Earlier attempts to manually generate attacks could not scale well and are limited by the attacker's expertise. However, the use of machine learning to automatically generate attacks has received attention~\cite{lin2019idsgan_IDSGAN}. Most of the work is focused on adversarial learning that attempts to generate  evasive attacks. In some cases\,\cite{zizzo2020adversarial_trustcom2020_Imperial} the researchers are able to generate stealthy attacks for a specific type of detector using the attack dataset from SWaT~\cite{sridhar_dataset_paper}.  Another work~\cite{kravchik2020poisoning_BGU_AsafShabtai} focused on the SWaT dataset to gradually poison the input samples during the training phase  to avoid detection during the test phase. They chose a set of seven attacks from the attack dataset and derived stealthy attacks for a threshold-based detector operating on the residual signal. These schemes make use of publicly available attack dataset~\cite{sridhar_dataset_paper} and modify those attacks to generate a limited number of stealthy attacks for the specific detection method. In~\cite{feng2017deep_3,lin2019idsgan_IDSGAN}, the authors used Generative Adversarial Networks (GANs) for learning anomaly detector classifiers and for the generation of malicious sensor measurements that will go undetected. A gradient-based adversarial attack scheme was proposed in \cite{JIA2021100452} which was able to deceive Recurrent Neural Network (RNN) based anomaly detector in two-real world CPS namely WADI\cite{wadi2017} and SWaT \cite{mathurTippenhauer}. Erba et al. ~\cite{erba_nils_2} has designed attacks to evade detection by an autoencoder in a water distribution systems~\cite{wadi2017}. They considered a white-box attacker that generates two different sets of spoofed sensor
values for the PLC and the detector. These studies focus on adversarial learning to deceive a specific classifier. Some techniques modify sensor data while others modify actuator data. In our method, we do impose any such limitations. Our proposed technique is an attempt to generate attack patterns autonomously using the limited attack dataset. 

Besides adversarial learning, there are techniques to generate synthetic data from a limited set of attack examples. A recent study uses the normal data from multiple sources and then generates anomalous data by sampling out of distribution data\,\cite{pham2014generating_attacks}. This method can generate anomalous data but that need not necessarily be the attack data, hence, not quite useful for security testing. There are a few well-known techniques to generate data samples, using the labeled examples from the minority class (anomalous data)~\cite{steinbusscalibration_SMOTE_others}. One such technique named  Synthetic Minority Over-sampling Technique (SMOTE)~\cite{chawla2002smote}, interpolates the current data samples to generate the artificial instances. However, though this oversampling method solves the imbalanced data problem for machine learning algorithms but does not provide any new scenarios but extends the provided minority examples. In contrast, the proposed technique provides thousands of new attack patterns, those never seen before and almost impossible to derive manually. \adityaIgnore{The goal of our technique is to learn the consequences/damage of the attack patterns designed by the process experts and learn all the different ways in which that kind of damage can be inflicted.}

\section{Conclusions}
A rule-based machine learning technique is proposed to autonomously generate attack patterns for the physical process of an ICS. More than 300,000 attack patterns were generated in a real-world six-stage water treatment plant. This is an incredible amount of attack scenarios as compared to $36$ public attack scenarios designed by domain experts on the same testbed. 
These insights are useful in testing the security of an ICS and providing additional data for signature-based intrusion detection systems. 
%=====================
% \bibliographystyle{IEEEtran}
% \bibliography{security.bib, references.bib}

% \begin{acks}
% This work was supported in part by the National Research Foundation (NRF), Prime Minister's Office, Singapore, under its National Cybersecurity R\&D Programme (Award No. NRF2016NCR-NCR002-023 and NRF2018NCR-NSOE005-0001) and administered by the National Cybersecurity R\&D Directorate.
% \end{acks}

%%
%% The next two lines define the bibliography style to be used, and
%% the bibliography file.
\bibliographystyle{ACM-Reference-Format}
\balance
\bibliography{references.bib}

\end{document}